# Credit Risk and Financial Performance of Commercial Banks: Evidence from Vietnam

Ha Nguyen[*]

**Abstract**

Credit risk is a crucial topic in the field of financial stability, especially at this time given the profound impact of the ongoing pandemic on the world economy. This study provides insight into the impact of credit risk on the financial performance of 26 commercial banks in Vietnam for the period 01/2006–05/2016. The financial performance of commercial banks is measured by return on assets (ROA), return on equity (ROE), and Net interest margin (NIM); credit risk is measured by the Non-performing loan ratio (NPLR); control variables are measured by bank-specific characteristics, including bank size (SIZE), loan loss provision ratio (LLPR), and capital adequacy ratio (CAR), and macroeconomic factors such as annual gross domestic product (GDP) growth and annual inflation rate (INF). The assumption tests show that models have autocorrelation, non-constant variance, and endogeneity. Hence, a dynamic Difference Generalized Method of Moments (dynamic Difference GMM) approach is employed to thoroughly address these problems. The empirical results show that the financial performance of commercial banks measured by ROE and NIM persists from one year to the next. Furthermore, SIZE and NPLR variables have a significant negative effect on ROA and ROE but not on NIM. There is no evidence found in support of the LLPR and CAR variables on models. The effect of GDP growth is statistically significant and positive on ROA, ROE, and NIM, whereas the INF is only found to have a significant positive impact on ROA and NIM.

[*] Department of Actuarial Studies and Business Analytics, Macquarie Business School, Macquarie University, Sydney, NSW 2109, AUSTRALIA; Email: ha.t.nguyen@mq.edu.au, ha.nguyen99804@gmail.com.

# 1 Introduction

Credit risk is most defined as the loss that results from the inability of a borrower to repay or meet the contractual obligations of a debt. The objective of credit risk management is to maximize a bank's risk-adjusted return by limiting credit risk exposure to acceptable levels. Banks must manage the inherent credit risk of their entire portfolio as well as the risk associated with individual credits or transactions. The effective management of credit risk is a crucial component of a comprehensive risk management strategy and essential to the long-term success of any financial institution.

Not only in Vietnam but in all countries in the world as well, banks are intermediate financial institutions that play a core role in the economy. Through their activities, banks contribute to ensuring the safety of the capital inflow and outflow of a country and bringing stability to the financial system. In the condition of a current market economy like Vietnam, the commercial banking system is a positive and key factor, and a financial intermediary between savings and investment, between organizations that have capital excess and a lack of funds. Especially after joining in the World Trade Organization on November 7, 2006, with the rapid growth of the economy, the competitive pressure in the banking sector became intense. Commercial banks have made a non-stop effort to expand the market as well as improve their profitability. Of these, credit activities are strongly pushed since they bring 80-90% of the income for each bank, but the risks are not less. Facing the opportunities and challenges in the process of international economic integration, the issue of raising the competitiveness of the domestic commercial banks with foreign commercial banks, especially improving the quality of credit, and minimizing risks has become urgent. Besides, in a complicated world economic situation and with the risk of the credit crisis, as a country with an open economy, Vietnam will not avoid the effects, especially at this time given the profound impact of the ongoing pandemic on the Vietnamese economy. This situation requires commercial banks in Vietnam to strengthen credit risk management to minimize risks affecting their profits.

Recognizing the importance of this issue, researchers conduct the world extensive studies on the impact of credit risk on the financial performance of commercial banks, such as Bourke (1989), Molyneux & Thorton (1992), Boahene (2012), Ongore & Kusa, 2013, and Noman et al. (2015). In Vietnam, there has been research on the impact of credit risk on the profitability of commercial banks (see, for example, Nguyen, 2019; Tran & Phan, 2020; and Phan et al., 2020). However, the works have not been thoroughly investigated and require further exploration, which creates space for this paper to be conducted.

This paper aims to seek empirical evidence to explain and discover how the credit risk has a relationship with the financial performance of Vietnamese Commercial Banks (VCBs). To obtain the objective, besides using the methods as Pooled OLS, FEM, REM, and FGLS, the dynamic Difference GMM approach is finally used to thoroughly address the problems. Empirical results show that the high significance of the lagged dependent variable's coefficients in the ROE and NIM models confirms the dynamic character of the model specification, thus justifying the use of dynamic panel data model estimation. The findings also report that despite the competition growth in Vietnamese financial markets, there is still a significant persistence of profit, measured by ROE and NIM, from one year to the next. It implies that if a bank makes an abnormal profit in the present year, then its expected profit for the following year will include a sizeable proportion of the present year's abnormal profit (Goddard et al., 2004). The empirical findings also suggest GDP has a significant positive impact on the financial performance of VCBs as measured by ROA, ROE, and NIM. On the contrary, beyond our expectations, LLPR and CAR are not found to be as important determinants of ROA, ROE, and NIM. It is also evident that the impact of SIZE, NPLR, and INF is not uniform across the different measures of VCBs' profitability used in this study. SIZE is found to be as an important determinant of profitability measured by ROA and ROE, but not by NIM. NPLR has a significant and negative impact on ROA and ROE, but the relationship is insignificant and negative in the NIM model. It is also evident that INF is positively and significantly related to ROA and NIM, but is insignificantly related to ROE.

The remainder of this study is organized in the following manner: Section 2 discusses the related literature on credit risk and the financial performance of the commercial banks. Section 3 describes the methodological approach. Section 4 provides the choice of variables and sample. Section 5 provides the findings and discussion of the result. Section 6 concludes the research work with useful policy implications.

## 2  Literature review and Hypothesis development

The risk appears as a possibility of unexpected outcomes in future events and hence exposes a financial institution to adversity (Emmett, 1997). In the literature on credit risk, modelling can be categorized into two large groups: structural and reduced form models. Structural models directly model the factors that affect the default process, such as the behaviour of underlying assets and the structure of cash flows. Reduced-form models start with the probability of changes in credit ratings or default using market data such as yields, recovery rates, and ratings change frequencies (see, for example, Jones, 2023, or Nguyen & Zhou, 2023 for an overview). In this paper, we focus on the impact of credit risk on the financial performance of commercial banks. In the banking system, the most fearful and critical risk is probably credit risk, which is the likelihood that a borrower will not be able to pay his debt on time or fail to make repayment at all (Sinkey, 2002). This is one of the most important areas of risk management and a matter of primary concern in the operation of banks since credit activities provide the main income for commercial banks. As a result, credit risk greatly affects the performance of banks, leading to banking institutions trying to develop their own credit risk models in order to improve bank portfolio quality. An increase in credit risk will result in higher marginal obligations and equity costs which will raise costs and expenses for the bank (Basel, 2004). In recent times, there have been lots of researchers interested in topics related to issues of credit risk and the performance of commercial banks. Two trends in the literature have concentrated on the major factors: internal factors (i.e., bank-specific characteristics) and external factors (macroeconomics), such as real GDP growth, inflation, and interest rates.

## 2.1 Bank performance indicators

For all financial institutions in general and commercial banks in particular, profit is the ultimate goal. All the strategies designed and their activities focus on this grand objective. Besides banks can have additional economic and social goals. This study is related to the first objective, profitability, which is an indicator of a bank's capacity to carry risk and/or increase its capital. In order to discover the impact of credit risk on the profitability of Vietnamese banks, the study uses return on assets (ROA), return on equity (ROE), and net interest margin (NIM) as profitability indicators, which are widely used in the literature (Tarusa *et al.*, 2012; Noman *et al.*, 2015).

## 2.2 Determinants of Bank Performance

There are mixed results from the extensive research about the impact of credit risk and the profitability of commercial banks. Abreu & Mendes (2001) conclude that the loan-to-asset ratio is significantly associated with the bank's profitability. They show that as higher deposits are transformed into loans, the profitability of commercial banks as measured by the net interest margin is higher. It implies that the bank will face more risk from having a higher loan ratio, thereby decreasing its profits. Boahene (2012) employs panel data to examine the relationship between credit risk and the six Ghanaian commercial banks from 2005-2009. He documents that credit risk is significantly positively associated with commercial bank profitability measured by ROE. Besides, Ruziqa (2012) investigates the effect of credit risk on the profitability of Indonesian commercial banks in the years 2007-2011. The results show that credit risk has a significant and negative effect on ROA, and ROE but is not significant on NIM. Recently, Noman *et al.* (2015) use unbalanced panel data from 18 commercial banks in Bangladesh from 2003 to 2013 to investigate the effect of credit risk on the profitability of the banks. Using POLS, REM, GLS, and system GMM, they find a robust negative and significant effect of non-performing loans, Loan Loss Provision ratio on ROA, ROE, and NIM. The analysis also finds a negative and significant effect of CAR on ROE.

The determinants of bank performance can be broken down into bank-specific factors that are controllable by management and macroeconomic factors that are beyond the control of the management of the bank. Each of these factors is further discussed below.

### 2.2.1 Bank-specific Factors/Internal Factors

As explained above, the internal factors are bank-specific characteristics that influence the profitability of the commercial bank. In accordance with the theoretical definition and literature review, these assorted potential variables for the Vietnam context in this study include Bank size, Non-performing loan ratio, Loan loss provision ratio, Capital adequacy ratio.

### 2.2.1.1 Non-performing loan ratio (NPLR)

NPLR occupies a significant portion of the total assets of insolvent banks and financial institutions (Fofack, 2005), and an increase in NPLR is a reflection of the failure of credit policy (Saba *et al.*, 2012). In recent studies, a large NPLR in the banking system is associated with bank failure and is a symptom of the economic slowdown as well (Khemraj & Pasha, 2012; Lata, 2014). Using the OLS, RE, GLS, and System GMM methods, Noman *et al*. (2015) find an inverse relation between the NPRL and the profitability of Bangladesh banks from 2003 to 2013. This is consistent with other research (Bourke, 1989; Molyneux & Thorton, 1992; Poudel, 2012; Kolapo *et al.*, 2012; Ruziqa, 2012). Therefore, we assert that increased credit risk weakens the performance of a bank.

**Hypothesis 1**. NPLRs are significantly and negatively related to the profitability of VCBs as measured by ROA, ROE, and NIM.

#### 2.2.1.2 Bank size (SIZE)

Bank size has a direct impact on profitability by reducing the cost of raising capital for large banks (Short, 1979). There are mixed impacts on the relationship between SIZE and the performance of commercial banks. Several studies by Vernon (1971), Smirlock (1985), Thornton (1992), Demirguc-Kunt & Maksimovic (1998), Demirgu9-Kunt & Huizinga (1999), Biker & Hu (2002), Fraker (2006), Pasiouras & Kosmidou (2007), and Gul *et al*. (2011) are in line with the theory that banks with bigger sizes tend to enjoy a higher level of profits. However, it is also evidenced that large banks show a significant negative relationship between size and profitability (Berger *et al*., 1987; Stiroh & Rumble, 2006). Batten & Vo (2019) find no correlation between the SIZE and the profitability measured by the NIM for Vietnamese commercial banks. The first hypothesis thus proposes that

**Hypothesis 2**. SIZE is positively and significantly related to the profitability of VCBs as measured by ROA, ROE, and NIM.

#### 2.2.1.3 Loan loss provision ratio (LLPR)

The use of LLPR to manipulate reported earnings has been widely discussed in the literature on banking and finance. Greenawalt & Sinkey (1988) describe that the banking industry tends to manipulate earnings more than other countries. A prior paper is illustrated by Scheiner (1981), who examines a sample of US commercial banks and concludes that LLPR is a vital tool used by bank management for managing earnings. Ma (1988) provides more evidence that during periods of high operating income, bank managers normally tend to raise the LLPR to lower the volatility of reported earnings. This outcome is supported by extensive studies for mainly US banks (Scholes *et al.*, 1990; Collins *et al.*, 1995; Liu *et al.*, 1997; and Ahmed *et al.*, 2014). Other empirical studies focusing on non-US banks also bring similar conclusions (Perez *et al.*, 2008; Kosmidou, 2008; Leventis *et al.*, 2011; and Norden & Stoian, 2014). Thus, we have the following hypothesis:

**Hypothesis 3**. LLPR is significantly and negatively related to the profitability of VCBs as measured by ROA, ROE, and NIM.

### 2.2.1.4 Capital adequacy ratio (CAR)

The safety of the banking system depends on the profitability and capital adequacy of banks. Several studies conclude that banks best perform when they maintain a high equity level relative to their assets (Demirgu9-Kunt & Huizinga, 1999; Goddard et al., 2004; Pasiouras & Kosmidou, 2007; Neceu & Goaied, 2008; Garcia-Herrero et al., 2009; Dang, 2011; Fungacova & Poghosyan, 2011; and Lee & Hsieh, 2013). They state that banks with higher capital ratios are likely to have lower funding costs due to lower prospective bankruptcy costs. Using OLS, RE, GLS, and, System GMM methods, Noman *et al.* (2015) investigate the relationship between CAR and the profitability of Bangladesh banks from 2003 to 2013. They find CAR is positively and significantly associated with the profitability of Bangladesh banks measured by ROA and NIM, which is consistent with the result of Kosmidou *et al.* (2005), while is negatively and significantly related to ROE (Choon *et al.*, 2012; Abiola & Olausi, 2014). We, therefore, assume

that

**Hypothesis 4**. CAR is significantly and positively related to the profitability of VCBs as measured by ROA, ROE, and NIM.

### 2.2.2 External Factors/ Macroeconomic Factors

External factors are not within the scope of the bank, and they indicate measures outside the impact of the bank. The macroeconomic factors and their effect on the profitability of companies have been discussed extensively, but results are not consensus.

#### 2.2.2.1 Gross Domestic Product Growth

A general agreement in the literature exists that GDP has an effect on the profitability of banks. This has been studied extensively, but the results lack consensus. Noman *et al.* (2015) use OLS, REM, GLS, and System GMM to investigate the effect of specific bank and macroeconomic factors on the profitability of 35 Bangladesh banks covering 2003 to 2013. The regression results indicate that GDP growth negatively affects profitability. This is consistent with the finding by Liu & Wilson (2010) but contradicts the findings from studies by Demirguc-Kunt & Huizinga (1999), Bikker & Hu (2002), Bashir & Hassan (2004), Pasiouras & Kosmidou (2007), Athanasoglou *et al.*, (2008), Omran (2011), Damena (2011), Saksonova & Solovjova (2011), and Bolt *et al.* (2012), who show that there is a positive impact of GDP growth on bank profitability. Therefore, we have the following hypothesis:

**Hypothesis 5**. GDP is statistically significantly related to the profitability of VCBs as measured by ROA, ROE, and NIM.

#### 2.2.2.2 Inflation Rate

Banking performance is highly influenced by inflation (INF). The findings from the previous studies regarding the relationship between inflation and profitability are varied. The studies of Barth *et al.* (1997), Demirguç-Kunt and Huizinga (1999), Denizer (2000), Abreu & Mendes (2001), and Khrawish (2011) show that there is a significant negative impact of inflation on banks' profitability. On the other hand, Abreu & Mendes (2001) use the Data-stream for the period of 1986-1999 for 477 banks from 4 different European countries (Portugal, Spain, France, and Germany) to conduct their research. The result acknowledges that the inflation rate brings along higher costs but also higher income. It means that bank costs are likely to increase more than bank revenues. This is consistent with earlier research conducted by Wallich (1977), Wallich (1980), Petersen (1986), Molyneux & Thornton (1992), Perry (1992), Guru *et al.*, (2002), Vong & Chan (2007), Tan & Floros (2012), and Frederic (2014). We, therefore, propose that

**Hypothesis 6**. INF rate is statistically significantly associated with the profitability of VCBs as measured by ROA, ROE, and NIM.

Based on the research objectives and discussion about the literature on credit risk and the financial performance of banks above, the study also seeks to test the following hypothesis

**Hypothesis 7**. The profitability of VCBs as measured by ROA, ROE, and NIM is persistent over time.

## 3 Methodology

Econometrics is widely used in research with various regression methods such as Pooled Ordinary Least Squares (POLS), Fixed Effects Model (FEM), Random Effects Model (REM), Feasible Generalized Least Squares (FGLS). Each method has its own strengths and drawbacks, including common problems such as multicollinearity, autocorrelation, or endogeneity. Once these assumptions are violated, then those methods are not efficient. Therefore, alternative methods for overcoming the detects are sought to employ, including the Generalized Method of

Moments (GMM).

## 3.1 Justifications of Dynamic Difference GMM

This paper employs Difference GMM to investigate the impact of credit risk on the financial performance of commercial banks. The GMM method was developed by Hansen (1982) as a non-parametric method for estimating model parameters. GMM is a desirable approach because it makes no distributional assumptions regarding the model specification. Consequently, GMM standard errors are resistant to autocorrelation and heteroskedasticity of an undetermined form. Under conditions of moderate regularity, the orthogonality conditions of the population model are specified, and the sample analogues are set to equal the population moments. The choice of instruments specified in the orthogonality conditions determines the necessary exogeneity assumption required for producing consistent parameter estimates. If an adequate instrument set is implemented in the orthogonality conditions, it is possible to generate estimates that are robust to simultaneity using the GMM estimation procedure. In addition, minor modifications to the equation of interest can eliminate the errors caused by unobservable heterogeneity and dynamic endogeneity.

Now suppose that we consider a model including $K$ parameters, $\boldsymbol{\theta} = (\theta_1, \theta_2, \dots, \theta_k)$ and that the theory provides a set of $L > K$ moment conditions,

$$E\left[\pi_l(x_i, y_i, z_i, \boldsymbol{\theta})\right] = E\left[\pi_{il}(\theta)\right] = 0,$$

where $x_i, y_i$ and $z_i$ are variables of the model; $i$ on $\pi_{il}(\theta)$ indicates the dependence on $(x_i, y_i, z_i)$. The corresponding sample means is denoted as

$$\bar{\pi}_l(X, Y, Z, \theta) = \frac{1}{n}\sum_{i=1}^{n}\pi_l\left(x_i, y_i, z_i, \theta\right) = \frac{1}{n}\sum_{i=1}^{n}\pi_{il}(\theta)$$

and then minimize this with respect to $\theta$. Unless the equations are functionally dependent, the system of $L$ equations in $K$ unknown parameters,

$$\bar{\pi}_l(\theta) = \frac{1}{n}\sum_{i=1}^{n} \pi_l(x_i, y_i, z_i, \theta) = 0, \qquad l = 1, \ldots, L,$$

will not have a unique solution. It will be necessary to reconcile the $\left(\frac{L}{K}\right)$ different sets of estimates which can be produced. The sum of squares can be applied to minimize a criterion function.

$$q = \sum_{l=1}^{L} \bar{\pi}_l^2 = \bar{\pi}(\theta)'\bar{\pi}(\theta)$$

or a weighted sum of squares,

$$q = \bar{\pi}(\theta)'W_n\bar{\pi}(\theta)$$

Holtz-Eakin et al. (1988) and Arellano and Bond (1991) developed the Difference GMM specification for dynamic panel datasets, which yields consistent parameter estimates in the presence of endogeneity. These estimates are robust to dynamic endogeneity, firm fixed-effects, endogenous regressors, heteroskedasticity, and serial correlation in firm performance innovation Schultz *et al.* (2010). Regarding the GMM method, Arellano and Bond (1991) proposed one- and two-step estimators. However, this study uses the one-step GMM estimator since Monte Carlo studies have found that this estimator outperforms the two-step estimator in terms of producing a smaller bias and a smaller standard deviation of the estimates (Judson & Owen, 1999; Kiviet, 1995).

In our case, to justify the most convincingly that dynamic Difference GMM[2] is employed as a valid tool in this study, we will examine Equation (2) below.

$$\pi_{it} = (\alpha + v_{it}) + \omega \pi_{it-1} + \sum_{k=1}^{6} \theta_k \xi_{it}^k + \varepsilon_{it}; \; u_{it} = v_{it} + \varepsilon_{it}, \quad (1)$$

where $\pi_{it}$: performance of commercial bank $i$ at time t measured by $ROA_{it}$, $ROE_{it}$, and $NIM_{it}$; $i$ represents the number of commercial banks, $i = 1...N$; $t$ represents the time, $t = 1... T$; $\alpha$: intercept; $\omega$ and $\theta$ are unknown coefficients; $\pi_{it-1}$ is one period lagged value of bank performance, $\xi_{it}$ is bank explanatory variables including SIZE, NPLR, LLPR, CAR, ΔGDP, ΔINF; $u_{it}$: disturbance term, $\varepsilon_{it}$: observation-specific errors, $v_{it}$: the unobserved specific effect.

Judson *et al.* (1996) show that the Fixed Effect (FE) model is biased when the time series *T* of panel data is small. Also, Nickell (1981) and Kiviet (1995) explain that regression coefficients will not be biased when *T* approaches infinity. It means that FE is efficient when the time series *T* is large enough. In other words, four major problems may arise when estimating Equation (2):

1. Explanatory variables can be considered endogeneity since causality may run in both directions - from the explanatory variables to dependent variables and vice versa. The regression of these variables can result in a correlation with the error term.

2. Fixed effects contained in the disturbance term in Equation (1) include characteristics of unobserved variables ($v_i$) and the observation-specific errors ($\varepsilon_{it}$)

$$u_{it} = v_i + \varepsilon_{it}$$

3. The presence of a lagged-dependent variable, $\pi_{it-1}$, in Equation (1) can give rise to autocorrelation.

---

[2] GMM includes the Difference and System Generalized Method of Moments (GMM) estimators. Using system GMM will increase efficiency, however, it uses more instruments than Difference GMM.

4. The panel data often has a short time dimension (small *T*) and a large individual dimension (large *N*).

To cope with problems (1) and (2), one should use Arellano-Bond (1991) Difference GMM as suggested by Holtz-Eakin, Newey & Rosen (1988). In dynamic Difference GMM, if the variables are predicted as endogenous, they would be categorized as variables instrumented under GMM-type; and only the lag of these variables are the appropriate instruments (Judson *et al.*, 1996). And if the explanatory variables are defined as strictly exogenous, they are also categorized as instrument variables (iv). For the strictly exogenous variables, their present and lagged values are the appropriate instruments (Judson *et al.*, 1996).

To resolve Problem (2), the Difference GMM uses the first difference to transform Equation (1) into:

$$\Delta \pi_{it} = \omega \Delta \pi_{it-1} + \sum_{k=1}^{6} \theta_k \xi_{it}^k + \Delta u_{it}, \qquad (2)$$

where $\Delta \pi$ is a $(M - N) \times 1$ vector of the differenced bank performance variable across *M* observations and *N* banks, $\omega$ is a $1 \times 1$ scalar of the coefficient for the lag time differenced performance measure $\Delta \pi_{it-1}$, across M observations; $\gamma$ is a $\pi \times 1$ vector of coefficients for the $\pi$ differenced bank explanatory variables; $\Delta \xi$ is a $(M - N) \times \pi$ matrix of the $\pi$ differenced bank explanatory variables across *M* observations and *N* banks, $\Delta u$ is a $(M - N) \times 1$ vector of error terms across *M* observations and *N* banks.

By transforming into the first difference regression, fixed effects will be excluded:

$$\Delta u_{it} = u_{it} - u_{it-1} = (v_i + \varepsilon_{it}) - (v_i + e_{it-1})$$

$$= (v_i - v_i) + (\varepsilon_{it} - \varepsilon_{it-1}) = \Delta \varepsilon_{it}$$

The first differenced lagged dependent variable is also instrumented with its past levels. Thus, problem (3) is also thoroughly processed. Finally, the Arellano-Bond Difference GMM estimator is designed for situations with small *T* and large *N* panel data (Problem 4) (Judson *et* *al.*, 1996). Therefore, the estimation by Arellano-Bond's (1991) one-step Difference GMM dynamic panel estimator is completely suitable for this study.

## 3.2 Reliability and Validations

Validations of Difference GMM are measured by two major assumptions concerning instrument variables. The rationality of instrument variables used in dynamic Difference GMM is assessed through Sargan statistics and Arellano-Bond (AR). Firstly, the Sargan test is to determine the suitability of instrument variables in GMM. This is a test for over-identifying the restrictions of the model. Sargan test states that $H_0$ of the instrument variables are exogenous, meaning that they are not correlated to disturbance of the model. Secondly, the Arellan-Bond test for autocorrelation has a $H_0$ of no autocorrelation and is applied to the differenced residuals. $H_0$ of the test for autocorrelation in first differences AR(1) is often rejected because $\Delta\varepsilon_{it} = \varepsilon_{it} - \varepsilon_{it-1}$ and $\Delta\varepsilon_{it-1} = \varepsilon_{it-1} - \varepsilon_{it-2}$ both have $\varepsilon_{it-1}$. Test for AR(2) in first differences is more important since it examines autocorrelation in levels.

## 4 Variables and Dataset

### 4.1 Dependent Variable

- ROA & ROE: are used as indicators of profitability in this study because they have been widely used in earlier research (see, for example, Naceur, 2003; Alkssim, 2005; Khrawish, 2011; Sehrish *et al*., 2011; Ongore & Kusa, 2013; Shumway, 2001; Nguyen, 2023).
- NIM is chosen as a dependent variable since it is often used in literature in recent years (see, for example, Rahman *et al.*, 2015; Khanh & Tra, 2015; and Batten & Vo, 2019).

### 4.2 Independent and Control Variables

Table 1 below contains the definitions of variables used as proxies of VCBs and of credit risk as well as their anticipated impacts according to the literature.

**Table 1**. Variables and definitions

| Variable | Definition | Measurement | Expected sign |
|---|---|---|---|
| $ROA_{it}$ | The ratio of return on assets for individual bank $i$ in time $t$. | $ROA_{it} = \dfrac{\text{Net profit after tax}_{it}}{(\text{Total Assets}_{t-1} + \text{Total Assets}_{t})/2} * 100\%$ | NA |
| $ROE_{it}$ | The ratio of return on equity for individual bank $i$ in time $t$. | $ROE_{it} = \dfrac{\text{Net profit after tax}_{it}}{(\text{Total equity}_{t-1} + \text{Total equity}_{t})/2} * 100\%$ | NA |
| $NIM_{it}$ | The ratio of net interest margin for individual bank $i$ in time $t$. | $NIM_{it} = \dfrac{\text{Net Interest Income}_{it}}{\text{Total Income Earnings Assets}_{it}} * 100\%$ | NA |
| $NPLR_{it}$ | The ratio of non-performing loans to total outstanding loans for individual bank $i$ in time $t$. | $LnNPLR_{it} = \dfrac{NPL_{it}}{\text{Gross loans}_{it}} * 100\%$ | - |
| $SIZE_{it}$ | The natural logarithm of the accounting value of the total assets of individual bank $i$ in time $t$. | $SIZE_{it} = Log(\text{Total Asset})$ | + |
| $LLPR_{it}$ | The ratio of provision for credit losses for loan customers to loans to customers for individual bank $i$ in time $t$. | $LLPR_{it} = \dfrac{\text{Provision for credit losses}_{it}}{\text{Total loans and advances}_{it}} * 100\%$ | - |
| $CAR_{it}$ | The ratio of own capital to risked-weighted assets for individual bank $i$ in time $t$. | $CAR_{i,t} = \dfrac{\text{Tier1 Capital}_{it} + \text{Tier2 Capital}_{it}}{\text{Total Risk weighted Asset}_{it}} * 100\%$ | + |
| $\Delta GDP_t$ | The annual growth rate in Gross Domestic Product in time $t$. | | +/- |
| $\Delta IFN_t$ | Annual inflation rate as measured by Consumer Price Index in time $t$. | | +/- |

*Notes*: All variables are computed by the author (except NPLR and CAR are collected from the report of the State Bank of Vietnam, and GDP &INF are taken from World Bank databases). –, + represent negative/positive impact, respectively. Independent variable includes NPLR; Control variables includes SIZE, LLPR, CAR, $\Delta GDP_t$, $\Delta IFN_t$

1. NPLR: is a major indicator of bank credit risk. This ratio is widely used in the literature on the financial performance of banks (see, for example, Poudel, 2012; Ruziqa, 2012).

2. SIZE: Total assets of banks are used as a proxy for bank size in the literature. Since the total assets of the bank in absolute terms are too large compared to the dependent variables in the model, it would not be appropriate to include total assets in its origin figures as an independent variable. Hence, this variable should be transformed by taking the natural logarithm of total assets before being included in the model (see, for example, Obamuyi, 2013; Frederic, 2014).

3. CAR: is heavily discussed in the literature on the financial performance of banks (see, for example, Obamuy, 2013; Ongore & Kusa, 2013; Frederic, 2014).

4. LLPR: Most studies discussed above come to the conclusion that bank managers are using this provision for different purposes, such as earning management and income smoothing. In this study, LLPR is used to identify the level of bank managers' expectations about asset quality of VCBs.

5. GDP growth: is used as a potential determinant of profitability. It is widely used in the literature (see, for example, Bikker & Hu, 2002; Athanasoglou *et al*., 2008; Rahman *et al.*, 2015).

6. INF rate indicates that the price level of the same general basket of goods and services increases over a time period. The change in price over time of the basket of goods and services is reflected by the consumer price index (CPI). This was heavily discussed in the literature (see, for example, Claessens *et al.*, 2001; Drakos, 2002; Alexiou & Sofoklis, 2009; Kasman *et al*., 2010; Tarusa *et al.*, 2012).

## 4.3 Endogenous, exogenous, and instrument variables

Endogenous variables are variables that are determined by models (Wooldridge, 2002), while exogenous variables are usually assumed to be determined by factors outside of models (Wooldridge, 2002). Regarding instrument variables, Arellano and Bond (1991) suggest that efficiency gains can be obtained by using available lagged values of the dependent variable and lagged values of the exogenous regressors as instruments. Based on the Hausman test, exogenous variables are chosen as instrument variables with suitable lagged values. Besides, as no-autocorrelation is correct, the lagged dependent variables are considered appropriate instrument variables (Wooldridge, 2001). Therefore, instruments are chosen for three models as follows: ROA Model: D.(L2.roa L3.gdp L.inf L.size nplr); ROE model: D.(L3.roe L2.gdp inf size nplr); and NIM model: D.(L2.nim L2.gdp size inf L.nplr L.car).

## 4.4 Data and sample

To investigate the relationship between credit risk and the financial performance of VCBs, 26 banks are selected, covering a 10-year period from 2006 to 2015 (Appendix 1). The data are collected from different sources, including the State Bank of Vietnam and individual banks' annual reports. Since the Vietnamese banking sector has developed in recent years, most banks have an average of 10 years of data. In Vietnam, on the 28$^{th}$ of June 2016, there are 30 commercial banks, however, five of them are excluded from the sample since the data is unavailable. Therefore, the sample size contains 255 observations at the time of conducting this study. Besides, macroeconomic variables including annual economic growth and the annual inflation rate, which are collected from the World Bank database, are also included to analyze this relationship.

## 5 Empirical Results

This section presents major results for investigating the relationship between credit risk and the performance of Vietnam's commercial banks. We also conduct some tests for the robustness of models and discuss the results using the dynamic Difference GMM approach.

## 5.1 Descriptive statistics

The descriptive statistics for all variables used in the models are summarized in the following table.

**Table 2.** Descriptive statistics of variables used in the model

| Variable | Mean | Std.Dev. | Min | Max |
|---|---|---|---|---|
| ROA | 1.13862 | 0.8076 | 0.01 | 5.54 |
| ROE | 10.5364 | 6.8449 | 0.07 | 35.25 |
| NIM | 2.97090 | 1.0934 | 0.44 | 7.36 |
| SIZE | 17.8207 | 1.3954 | 13.57 | 20.59 |
| NPLR | 2.1610 | 1.4431 | 0.00 | 11.40 |
| LLPR | 1.2865 | 0.8476 | 0.01 | 5.26 |
| CAR | 14.7724 | 6.4919 | 4.82 | 54.92 |
| GDP | 6.09843 | 0.6485 | 5.20 | 7.10 |
| INF | 9.39058 | 6.3007 | 0.60 | 23.10 |

As can be seen from Table 2, SIZE observes a range from 13.57 to 20.59, with 17.82 on average. NPLR has a minimum value of 0% with VIB and a maximum value of 11.4 % with SCB. The maximum NPLR is 11.4%, which is higher than the regulatory requirement of 3%. This reflects weak credit risk management by VCBs. LLPR varies from 0.01% for MBBank to 5.26% for BIDV. CAR reaches the highest value of 54.92% by VietCapitalBank and has the lowest value of 4.82% by BIDV, with a mean and standard deviation of 14.77% and 6.49%, respectively, which indicates that there is high volatility among the banks' ability to manage risk management. The minimum CAR is 6.78%, which is lower than the regulatory requirement of 9%, which is evidence of the non-compliance of some VCBs regarding Basel II requirements[3].

NIM fluctuates from 0.44% to 7.36%, with an average of 2.97%. ROE reaches a maximum value of 35.25%, drops to a minimum level of 0.07%, and takes an average of 6.84%. ROA reaches its highest value, at 5.54%, and its lowest value, about 0.01%, with an average and standard deviation of 1.14% and 0.81%, respectively. In general, banks are competing amongst themselves to make a profit; however, their standard deviations show that their profit-making capacity is different from each other.

---

[3] SBV publicizes Circular No.13/2010/TT-NHNN stipulating a CAR for VCBs of 9% between their total own capital and total risk-weighted assets under Basel Accord (II), which is a set of agreements for commercial banks.

## 5.2 Correlation Analysis

**Table 3.** Correlation Matrix for variables

| Variables | ROA | ROE | NIM | SIZE | NPLR | LLPR | CAR | GDP | INF |
|---|---|---|---|---|---|---|---|---|---|
| ROA | 1.0000 | | | | | | | | |
| ROE | 0.6069 | 1.0000 | | | | | | | |
| NIM | 0.3045 | 0.0863 | 1.0000 | | | | | | |
| SIZE | -0.4191 | 0.1241 | -0.0952 | 1.0000 | | | | | |
| NPLR | -0.3359 | -0.3667 | 0.0123 | 0.1402 | 1.0000 | | | | |
| LLPR | -0.2977 | -0.0802 | -0.0026 | 0.5039 | 0.4752 | 1.0000 | | | |
| CAR | 0.1981 | -0.1832 | 0.2460 | -0.5078 | 0.0513 | -0.2289 | 1.0000 | | |
| GDP | 0.2708 | 0.2489 | -0.1461 | -0.2134 | -0.3397 | -0.2316 | 0.0143 | 1.0000 | |
| INF | 0.1617 | 0.1323 | 0.1450 | -0.1525 | 0.0618 | 0.0184 | 0.0702 | -0.1983 | 1.0000 |

Table 3 shows the correlation between the independent variables with respect to ROA, ROE, and NIM. ROA is positively affected by CAR, GDP, and INF while is negatively associated with SIZE, NPLR, and LLPR. Also, ROE is positively associated with SIZE, GDP, and INF, whereas it is negatively associated with NPLR, LLPR, and CAR. Finally, NIM is positively associated with NPLR, CAR, and INF, whereas it is negatively correlated with SIZE, NPLR, and GDP. On the other hand, according to Kennedy (2008), the model tends to have multicollinearity when the correlation exceeds 0.80. Hence, there is no existence of multicollinearity in this case. This is also firmly affirmed in multicollinearity test results with Variance Inflation Factor-VIF[4] (Appendix 2). Besides, the matrix also shows three pairs of high-correlation variables, including SIZE and LLPR (0.51), SIZE and CAR (-0.51), and NPLR and LLPR (0.48) which are likely to be endogenous[5] variables. Therefore, these variables will be examined later with the Hausman and Sargan tests.

## 5.3 Model assumptions

This part conducts some tests of model assumptions: (i) Unit root test (stationarity), ii)

---

[4] See Gujarati (2022, p. 340). Multicollinearity exists in the model if VIF > 10.0. The VIF for the variables in the model ranges from 1.10 to 1.85, suggesting the absence of multicollinearity.
[5] See Wooldridge J. (2009, p. 553). Endogeneity problems in our model can be due to omitted variables and measurement error.

Autocorrelation, (iii) Heteroskedasticity, and (iv) Exogeneity and Endogeneity. Firstly, Fisher's test[6] using the four methods proposed by Choi (2001) is employed to test stationarity (unit root) since it is flexible; it does not require strongly balanced data, and the individual series can have gaps. The test result indicates that under all four methods of the tests, $H_0$ of the presence of unit root is robustly rejected (p is close to 0). It implies that all variables in the models are stationary (see Appendix 3). Secondly, we use the Wooldridge test to test for autocorrelation.

**Table 4.** Autocorrelation tests

| Method | Variables | F-statistic | Prob > F | Result |
|---|---|---|---|---|
| Wooldridge test | ROA | 2.506 | 0.1260 | No autocorrelation |
|  | ROE | 17.430 | 0.0003 | Autocorrelation |
|  | NIM | 48.377 | 0.0000 | Autocorrelation |

**Notes**: Under the Wooldridge test, $H_0$ indicates no autocorrelation

Table 4 shows that the p-value of ROA is 0.1260, which is higher than the 10% level, meaning that it is impossible to reject $H_0$ and conclude that ROA has no autocorrelation, whereas ROE and NIM have a small p-value (close to 0), implying that autocorrelation exists in both of these models.

Thirdly, we use the Breuch-Pagan, Wald, and Breuch & Pagan Lagrangian Multiplier tests of Heteroskedasticity.

---

[6] Under Fisher-test (Choice, 2001), $H_0$ is all the panels containing unit roots (i.e., variables of time-series are non-stationary).

**Table 5.** Heteroskedasticity tests

| Models | Tests | Variables | Chi2 | Prob>Chi2 | Result |
|---|---|---|---|---|---|
| POLS | Breuch-Pagan | ROA | 17.65 | 0.0000 | Heteroskedasticity |
| | | ROE | 5.56 | 0.0184 | |
| | | NIM | 16.81 | 0.0000 | |
| FEM | Wald | ROA | 1378.02 | 0.0000 | Heteroskedasticity |
| | | ROE | 451.30 | 0.0000 | |
| | | NIM | 463.91 | 0.0000 | |
| REM | Breuch & Pagan Lagrangian Multiplier | ROA | 38.35 | 0.0000 | Heteroskedasticity |
| | | ROE | 37.93 | 0.0000 | |
| | | NIM | 172.58 | 0.0000 | |

**Notes**: The Breuch-Pagan, Wald, and Breuch & Pagan Lagrangian Multiplier tests have $H_0$ of constant variance (homoskedasticity).

From Table 5, under Pooled OLS, the Breusch-Pagan test is used and gets results with small p-values (<5%), which means that $H_0$ is rejected. Hence, all three models violate the assumption of OLS. Under FEM, the Wald test is employed, and $H_0$ of the three models is robustly rejected with small p-values (close to 0). A similar result is obtained for REM using the Breuch & Pagan Lagrangian Multiplier test. In sum, three models have non-constant variances.

Finally, using Durbin and Wu-Hausman to test for endogenous variables, the test results (see Appendix 4) show that all variables in the three models are exogenous, excluding the LLPR variable, since the p-values of LLPR in the three models are smaller than the 1% level in the ROA and NIM models and the 5% level in the ROE. This indicates that $H_0$ of exogeneity is rejected and concludes that the LLPR is endogenous.

### 5.4 Estimation Results and Discussion

From the basic panel data, we will conduct the estimation for the model with Pooled

OLS, FE, RE, and FGLS methods. The purpose of this is to test the sensitivity and the structural changes of variables in the models, thereby discovering the preliminary phase about the effect of these variables as well as evaluating the robustness of models. However, these methods give weak estimation results since they do not cope with the endogeneity of explanatory variables. From the dynamic panel data in Equation (2), the Difference GMM method, which can overcome the defects, is employed to estimate models.

### 5.4.1 Pooled OLS, FEM, and REM estimators

Table 6 shows the basic estimation results under methods of Pooled OLS, FE, and RE. Overall, the fit of the model is very good (F-tests and Wald chi2 for models are statistically significant at a 1% level). Based on three pairs of specification test results (Pooled OLS against FEM, Pooled OLS against REM, and FEM against REM) in Table 6, it can be seen that there is a cross-section fixed effect and fixed period in the ROA and ROE models, whereas there is a cross-section random effect in the NIM model. Therefore, FEM is preferable for estimating ROA and ROE, while REM is preferable for estimating NIM. The result of the estimation is as follows: *Firstly,* under the FE method, bank size is significant in the ROA and ROE models at 1% and 5% levels, respectively. NPLR is significant and negative at 5% and 1% levels in ROA and ROE, respectively. LLPR is significant at a 5% level in ROE but is insignificant in ROA. CAR is insignificant in both ROA and ROE. In respect of macroeconomic variables, Table 6 also reports that GDP and INF are significant in the ROE model but are insignificant in the ROA model. *Secondly*, under the RE method, SIZE, NPLR, and LLPR are insignificant in the NIM model. On the contrary, CAR, GDP, and INF are significant at the 5% level.

**Table 6.** Pooled OLS, FEM, and REM estimator results

| Models | Pooled OLS | | | FEM | | | REM | | |
|---|---|---|---|---|---|---|---|---|---|
| Variables | ROA | ROE | NIM | ROA | ROE | NIM | ROA | ROE | NIM |
| SIZE | -.1853557*** (-4.39) | 1.070575*** (2.92) | .0029316 (0.05) | -.4123288*** (-7.19) | -1.043944** (-2.12) | .0363617 (0.47) | -.2686018*** (-5.74) | .2262472 (0.55) | .0100103 (0.14) |
| NPLR | -.1497774*** (-4.14) | -1.557273*** (-4.95) | -.0648966 (-1.18) | -.0906411** (-2.62) | -.8884011*** (-2.98) | -.0007577 (-0.02) | -.1216472*** (-3.52) | -1.191662*** (-3.99) | -.0070416 (-0.15) |
| LLPR | .0310078 (0.46) | -.0181262 (-0.03) | .0784962 (0.76) | -.0494488 (-0.66) | -1.341707** (-2.09) | .0945287 (0.94) | .0048607 (0.07) | -.6415321 (-1.05) | .0825829 (0.86) |
| CAR | .0054285 (0.69) | -.0800760 (-1.17) | .0439707*** (3.68) | .003312 (0.41) | -.074253 (-1.07) | .0170009 (1.56) | .0009291 (0.12) | -.1070731 (-1.58) | .0221401** (2.09) |
| GDP | .1859535** (2.51) | 2.443288*** (3.79) | -.242164** (-2.14) | .0933476 (1.35) | 1.37424** (2.31) | -.1570313* (-1.69) | .1618299** (2.36) | 2.070909*** (3.51) | -.1821264** (-2.00) |
| INF | .0199197*** (2.77) | .2576353*** (4.12) | .0178579 (1.63) | .010581 (1.61) | .1607981*** (2.85) | .0224415** (2.53) | .0170102** (2.59) | .2218399*** (3.92) | .0206702** (2.38) |
| _CONS | 3.324309*** (3.27) | -21.29009** -2.41 | 3.617485** (2.33) | 8.028566*** 6.30 | 23.99251** (2.19) | 2.698696 (1.57) | 5.026365*** (4.67) | -3.22894 (-0.34) | 3.296108** (2.11) |
| | | | | **Cross-section fixed Period fixed** | | | **Cross-section random effect** | | |
| F-statistics Prob>F | 16.77 0.0000 | 13.82 0.0000 | 4.52 0.0002 | 22.90 0.0000 | 13.36 0.0000 | 3.50 0.0025 | | | |
| Wald chi2 Prob>F | | | | | | | 118.36 0.0000 | 71.88 0.0000 | 22.26 0.0026 |

**Specification Test for Models**

| Tests | F-test for Pooled OLS against FEM | | | Breusch & Pagan Lagrangian test for Pooled OLS against REM | | | Hausman test for FEM against REM | | |
|---|---|---|---|---|---|---|---|---|---|
| | ROA | ROE | NIM | ROA | ROE | NIM | ROA | ROE | NIM |
| | F (25,223) = 4.42 Prob>F = 0.000 | F (25,223) = 4.69 Prob>F = 0.0000 | F (25,223) = 8.05 Prob>F = 0.0000 | Chibar2(01) = 38.35 Prob>Chiba2 = 0.000 | Chibar2(01) = 37.93 Prob>Chiba2 = 0.0000 | Chibar2(01) = 172.58 Prob>Chiba2 = 0.000 | Chi2(6) = 28.65 Prob>chi2 = 0.0001 | Chi2(6) = 36.08 Prob>chi2 = 0.0000 | Chi2(6) = 5.66 Prob>chi2 = 0.4626 |
| Result | **Fixed Effect** | | | **Random effect** | | | **Fixed Effect** | **Fixed Effect** | **Random Effect** |

**Note**: ***, **, and * denote significance at 1%, 5%, and 10%, respectively; H$_0$ of the F-test for Pooled OLS against FEM: there is no heterogeneity (FE is preferred). H$_0$ of Breusch & P. L. for Pooled OLS against REM: there is no heterogeneity (RE is preferred). H$_0$ of Hausman test: no correlation between regressors and random effects (REM is preferred) and in case there is a correlation between regressors and random effects (FEM is preferred). With a p-value <1%, H$_0$ will be rejected. Under F-test and Breusch & P.L test, the p-value <1%. As a result, under F-test, FEM is preferred in three models, under Breusch & P.L, REM is correct in three models; then the Hausman test is used to select between FEM and REM. Therefore, as can be seen, ROA and ROE have cross-section fixed Effects and fixed period (p-value< 1%, H$_0$ is rejected), so FEM is appropriate, and NIM has random effects (p-value = 0.46>10%), so REM is preferred.

### 5.4.2 Feasible Generalized Least Squares (FGLS) estimator

As analyzed above, based on specification test results, we have determined that ROA and ROE have the cross-section fixed effect and fixed period, whereas NIM has the cross-section random effect. However, since the model in this study has severe autocorrelation in the ROE and NIM models and heteroskedasticity in all three models, the estimation results from FEM and REM are not reliable; thus, the FGLS estimator is used to solve these problems.

**Table 7**. Comparison results between FGLS and FEM & REM estimator

| Variables | ROA | ROE | NIM | ROA | ROE | NIM |
|---|---|---|---|---|---|---|
| SIZE | -.1564812*** | .8863494*** | -.0299748 | -.4123288*** | -1.043944** | .0100103 |
|  | (-4.99) | (2.81) | (-0.45) | (-7.19) | (-2.12) | (0.14) |
| NPLR | -.1531227*** | -1.430926*** | -.038074 | -.0906411** | -.8884011*** | -.0070416 |
|  | (-6.74) | (-5.43) | (-0.91) | (-2.62) | (-2.98) | (-0.15) |
| LLPR | .0196721 | .289889 | -.0324919 | -.0494488 | -1.341707** | .0825829 |
|  | (0.54) | (0.60) | (-0.43) | (-0.66) | (-2.09) | (0.86) |
| CAR | .0064894 | -.044829 | .0182932* | .003312 | -.074253 | .0221401** |
|  | (0.92) | (-0.83) | (1.72) | (0.41) | (-1.07) | (2.09) |
| GDP | .1131847** | 1.542823*** | -.1512276** | .0933476 | 1.37424** | -.1821264** |
|  | (2.21) | (3.30) | (-2.02) | (1.35) | (2.31) | (-2.00) |
| INF | .0170661*** | .0076262 | .0247401*** | .010581 | .1607981*** | .0206702** |
|  | (3.41) | (0.18) | (3.76) | (1.61) | (2.85) | (2.38) |
| _Con | 3.259723*** | -12.06387* | 2.154567* | 8.028566*** | 23.99251** | 3.296108** |
|  | (4.24) | (-1.76) | (2.99) | 6.30 | (2.19) | (2.11) |
| Wald chi2(6) | 186.11 | 70.79 | 26.44 |  |  | 22.26 |
| Prob>chi2 | 0.0000 | 0.0000 | 0.0002 |  |  | 0.0011 |
| F-test |  |  |  | 22.90 | 13.36 |  |
| Prob>F |  |  |  | 0.0000 | 0.0000 |  |
| Estimator | FGLS | | | FEM | | REM |

**Notes**: ***, **, and * denote significance at 1%, 5%, and 10% levels respectively, t-statistics are shown in parentheses.

Table 7 reports the goodness of fit in all three models, which is evident from the significant Wald test and F-test at a 1% level in all models. Based on the results of three pairs of specification tests for the three models above (POLS against FEM, POLS against REM, and FEM against REM), obviously there is a fixed cross-section effect on the ROA and ROE models and a random cross-section effect. On that basis, we will compare results from FGLS in ROA, ROE, and NIM with those from the FE method in ROA and ROE model and the RE method in the NIM model. As observed, SIZE and NPLR are rather persistent under the FGLS and FE methods, except for a minor change is that under the FGLS method, NPLR is significant at a 1% level in the ROE model, while under FEM it is significant at a 5% level in the ROE model. LLPR and CAR are significant at 5% and 10% respectively in ROE under FEM, but both of them are insignificant under FGLS. Conversely, GDP becomes significant at 5% and 1% levels in ROA and ROE, respectively, under FGLS. Similarly, under FGLS, INF is significant at a 1% level in ROA but insignificant in the ROE model. On the other hand, in the NIM model, results are rather consistent under both methods, with just a minor change: CAR becomes less significant while INF becomes more significant under FGLS.

However, when resolving autocorrelation and non-constant variance, the two following problems should be considered under FGLS. *Firs*t, FGLS requires strict exogeneity assumptions on the regressors (Wooldridge, 2001). *Second*, according to Wooldridge (2016, p. 384), one should be careful with small sample sizes under FGLS. In this study, the sample size is categorized into medium and small groups, so to guarantee more reliable and efficient estimation results, dynamic Difference GMM estimation will be employed later as a final regression model to analyze and discuss this study.

### 5.4.3 Difference Generalized Method of Moments (Difference GMM) estimator

This part provides the empirical results based on dynamic Difference GMM method. Assumption tests for the validity of the Difference GMM are also conducted to check the robustness of the model. As mentioned above the Dynamic Difference GMM is used as the final method to analyze and discuss the impact of credit risk on the performance of VCBs.

**Table 8.** Dynamic Difference GMM estimator results

| Variables | Model 1 (ROA) | Model 2 (ROE) | Model 3 (NIM) |
|---|---|---|---|
| L1.(ROA,ROE, NIM) | .0319656 (0.21) | .4794182** (2.37) | 1.283483*** (3.17) |
| SIZE | -.8871043 *** (-5.62) | -5.1484*** (-3.78) | -.3189178 (-1.02) |
| NPLR | -.0763054* (-1.67) | -.9785491* (-1.88) | -.2633495 (-0.92) |
| LLPR | -.0523489 (-0.24) | 3.050213 (1.26) | -.3037985 (-0.66) |
| CAR | -.0030461 (-0.14) | -.3849225 (-1.10) | .0398092 (0.65) |
| GDP | .1910464** (2.33) | 2.160597** (2.54) | .5929881* (1.71) |
| INF | .024313** (2.41) | -.0937191 (-0.75) | .0543595*** (2.89) |
| No. of group | 26 | 26 | 26 |
| No of instruments | 26 | 26 | 13 |
| Instrument variables (iv) | D.(L2.roa L3.gdp L.inf L.size nplr) | D.(L3.roe L2.gdp inf size nplr) | D.(L2.nim L2.gdp size inf L.nplr L.car) |
| GMM-type | L(1/.).L3.llpr | L(1/.).L3.llpr | L.L.llpr |
| Arellano-Bond test for AR (1) | z = -3.01 Pr>z = 0.003 | z = -2.81 Pr>z = 0.005 | z = -3.90 Pr>z =0.000 |
| Arellano-Bond test for AR (2) | z = 1.43 Pr>z = 0.153 | z = 1.64 Pr>z = 0.101 | z = -0.97 Pr>z = 0.330 |
| Sargan test of overid.restriction | chi2(19) = 16.69 prob>chi2 = 0.611 | chi2(19) = 14.13 prob>chi2 = 0.776 | chi2(6) = 5.74 prob>chi2 = 0.453 |
| Difference-in-Hansen tests of exogeneity of instrument subsets | | | |
| Sargan test excluding group | Chi2(14) = 10.80 Prob>chi2 = 0.702 | Chi2(14) = 10.09 Prob>chi2 = 0.756 | Chi2(0) = 0.00 Prob>chi2 = 1.000 |
| Difference (Ho = exogenous) | Chi2(5) = 5.89 Prob>chi2 = 0.317 | Chi2(5) = 4.05 Prob>chi2 = 0.543 | Chi2(6) = 5.74 Prob>chi2 = 0.453 |
| Estimation Method | Dynamic Difference GMM | Dynamic Difference GMM | Dynamic Difference GMM |

**Notes**: ***, **, and * denote significance at 1%, 5%, and 10% levels, respectively; with a 95% confidence interval. The estimation method is Arellano & Bond's (1991) one-step GMM dynamic panel estimator. Arellano-Bond test that autocorrelation in residuals of order 1 and order 2 is equal to 0 (H$_0$: No autocorrelation). Under the Sargan test for the validity of over-identifying restrictions, the H$_0$ of instrument variables is valid. P-value <0, H$_0$ will be rejected.

From Table 8, it can be seen that dynamic Difference GMM outputs indicate the *goodness of fit of the models,* which is evident from the instrument variables which are smaller than or equal to the number of groups and the high significance of AR(2) and Sargan test. In a comparison between FGLS and Difference GMM estimates in Tables 7 and 8, respectively, one notes that the results produced by the two methods in all three models are similar, meaning that the estimation results tend to be consistent and accurate.

As mentioned above, the grand objective of this study is to discover the impact of credit risk on VCBs' profitability. Hence, seven different hypotheses are established to investigate this relationship. In this study, the analysis of regression results by dynamic Difference GMM approach is used along with the correlation analysis to test the hypotheses; the test results for these hypotheses are presented as follows:

**Hypothesis 1**: *NPLRs are significantly and negatively related to the profitability of VCBs as measured by ROA, ROE, and NIM.*

Table 8 shows that the coefficients of NPLR of -0.08 and -0.98 with p-values of 0.098 and 0.062 in the ROA and ROE models imply that NPLR is weakly significant at the 10% level. The significant and negative coefficients of NPLR indicate that an increase of 1% in NPLR will reduce VCBs' profitability as measured by ROA and ROE by 0.08% and 0.98%, respectively. The result of the ROA model is in agreement with plenty of the previous empirical research by Poudel (2012), Kolapo *et al.* (2012), Boahene *et al.* (2012), Ruziqa (2012), and Noman *et al.* (2015). Also, the result where ROE is the dependent variable corroborates the findings of Bourke (1989), Molyneux & Thorton (1992), Ruziqa (2012), and Noman *et al.* (2015). Besides, the coefficient of NPLR in the NIM model is negative but insignificant. This result is consistent with Ruziqa (2012) but contradicts the findings of Khanh & Tra (2015) and Rahman *et al.* (2015), suggesting that the NPLR is positive and insignificant on NIM, and also contradicts the finding of Noman *et al.* (2015), who reveal a negative and significant effect of NPLR on NIM.

Overall, the results of ROA and ROE are in line with our prior expectations. With regard to the hypothesis tested, Hypothesis 2 is valid where ROA and ROE are dependent variables since NPLR has a significant impact on the VCBs' profitability, while Hypothesis 2 is invalid where the VCBs' profitability is measured by NIM due to the insignificant coefficient of NPLR.

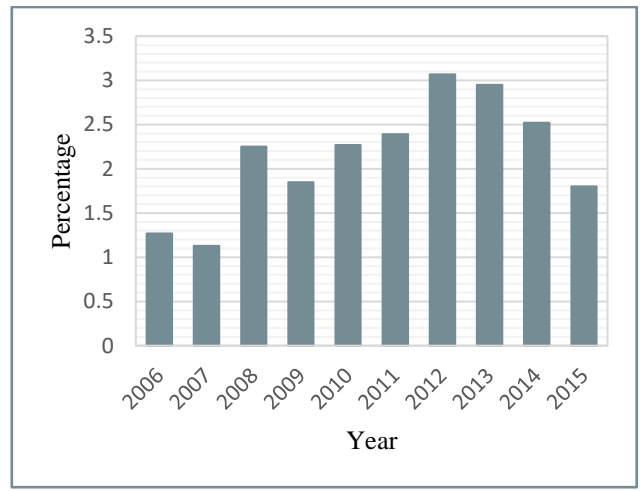

**Figure 1**: Non-performing loans of VCB (2006-2015). *Source*: Annual Report of VCBs

A positive signal is that the trend on NPLR in VCBs showed a decline (Figure 1), indicating that managers and policymakers in Vietnam have enhanced credit risk management mechanisms in the banking industry.

**Hypothesis 2**: *SIZE is positively and significantly related to the profitability of VCBs as measured by ROA, ROE, and NIM.*

Bank size has a significant negative effect on profitability measured by ROA and ROE. The result in Table 8 shows that an increase of one unit in SIZE will reduce ROA and ROE by 0.89% and 5.15%, respectively. The result of ROA is in consonance with Berger *et al.* (1987), Stiroh & Rumble (2006), and Ruziqa (2012), while contradicting the findings of Molyneux & Thornton (1992), Bikker & Hu (2002), Goddard *et al.* (2004), Gul *et al.* (2011), and Rahman *et* *al.* (2015), who reveal that SIZE has a significant and positive effect on ROA. Ruziqa (2012) and Frederic (2014) do not also support this finding, suggesting that SIZE has a positive but insignificant relation with ROA. In the case where profitability is measured by ROE, this result contradicts the findings of Ruziqa (2012), suggesting that SIZE has no impact on ROE.

On the other hand, the result where NIM is the dependent variable shows that the coefficient of SIZE is negative but insignificant. This finding supports the studies conducted by Rahman *et al.* (2015) and Batten & Vo (2019), while contradicting the results of Berger *et al.* (1987), Stiroh & Rumble (2006), and Ruziqa (2012), suggesting that SIZE has a negative and significant impact on NIM.

With regard to the hypotheses tested, we can conclude that Hypothesis 1 is valid where ROA and ROE are dependent variables since SIZE has a significant impact on the VCBs' profitability, while Hypothesis 1 is invalid where the VCBs' profitability is proxied by NIM due to the insignificant coefficient of SIZE. However, SIZE did not affect ROA and ROE in the same direction as expected. According to data of VCBs, the SIZE of has increased year by year but has not benefited from economic scale, i.e., the larger VCBs become, the less efficiently they work. It is not surprising that SIZE has a negative relationship with VCBs' profitability, which is consistent with Berger & DeYoung's (1997) skimping hypothesis, which states that banks maximizing long-term profits can opt to have lower costs in the short-term run by skimping on resources devoted to tracking loans, but carry the consequences of greater loan activity issues.

**Hypothesis 3**: *LLPR is significantly and negatively related to the profitability of VCBs as measured by ROA, ROE, and NIM.*

Beyond our expectation, the credit risk proxied by LLPR is insignificant in the ROA,

ROE, and NIM models. For profitability measured by ROA, this result is consistent with Rahman et al. (2015) but contradicts Anandarajan et al. (2003), who conclude that there is a significant and positive relationship between LLPR and banks' profitability. This also contradicts the findings of Trujillo-Ponce (2012), Ongore & Kusu (2013), and Samina & Ayub (2013), who discovered that LLPR has a significant negative impact on banks' profitability measured by ROA. On the other hand, when profitability is proxied by ROE and NIM, the result contradicts the findings of Tarusa et al. (2012), Kolapo et al. (2012), Rahman et al. (2015), and Noman et al. (2015), who document that LLPR is negatively and significantly associated with ROE and NIM. Regarding Hypothesis 3, since LLPR has an insignificant impact on the VCBs' profitability, LLPR is not found to be an important determinant of ROA, ROE, and NIM; implying that Hypothesis 3 is rejected.

**Hypothesis 4**: *CAR is significantly and positively related to the profitability of VCBs as measured by ROA, ROE, and NIM.*

Similar to LLPR, beyond our expectation, the coefficient of CAR is negative in the ROA and ROE models while it is positive in the NIM model but is statistically insignificant. This implies that CAR has no impact on the VCBs' profitability. The result where ROA is the dependent variable is in line with Qin & Dickson (2012), whereas contradicts other previous researchers (Ongore & Kusa, 2013; Obamuy, 2013; Frederic, 2014), who describe that CAR has a highly negative impact on ROA, and the finding of Rahman et al. (2015), who find a positive and significant impact on ROA. The results, where ROE and NIM are dependent variables, contradict Noman et al. (2015) and Rahman et al. (2015), who discover a significant effect of CAR on ROE and NIM. Thus, Hypothesis 4 is rejected because CAR has an insignificant impact on the VCBs' performance.

**Hypothesis 5**: *GDP is statistically significantly related to the profitability of VCBs as measured by ROA, ROE, and NIM.*

On the side of external factors, as seen in Table 8, GDP has significant and positive coefficient estimates at a level of 5% in the ROA and ROE models and at a level of 10% in the NIM model. The results show that an increase of 1% in GDP will increase the VCBs' profitability, proxied by ROA, ROE, and NIM, by 0.19%, 2.16%, and 0.59%, respectively. The results of the impact of GDP growth on ROA, ROE, and NIM are consistent with Pasiouras *et al.* (2007), Kosmidou (2008), Rahman *et al.* (2015), and Batten & Vo (2019). In the case where ROE is used to measure the VCBs' profitability, the result is consistent with Batten & Vo (2019) but contradicts Rahman *et al.* (2015), who find a positive but insignificant association between ROE and GDP. A positive and significant correlation between GDP and NIM is also supported by the finding of Batten and Vo (2019), whereas Rahman *et al.* (2015), suggest no such significant correlation. Therefore, Hypothesis 5 is valid, indicating that annual GDP growth plays an important role in determining the VCBs' profitability.

**Hypothesis 6**: *INF is statistically significantly associated with the profitability of VCBs as measured by ROA, ROE, and NIM.*

The regression results in Table 8 show that the inflation rate in the ROA and NIM models has coefficients of 0.02 and 0.05 and p-values of 0.017 and 0.004, respectively, meaning that INF has a significant and positive impact at a 5% level on ROA and a 1% level on NIM within a 95% confidence interval. The results indicate that an increase of 1% in INF would increase by 0.02% and 0.05% respectively of the VCBs' profitability proxied by ROA and NIM. These findings contradict Rahman *et al.* (2015) while being consistent with other extensive studies conducted by Demirguf-Kunt & Huizinga (1999), Brock & Suarez (2000), Claessens *et* *al.* (2001), Drakos (2002), Alexiou & Sofoklis (2009), Kasman *et al.* (2010), Tarusa *et al.* (2012), and Khanh &Tra (2015). Table 8 also reports that INF is negatively associated with ROE, but the relationship is insignificant. The finding is inconsistent with the result of Rahman *et al.* (2015), who suggest that INF is negatively and significantly associated with ROE.

In sum, Hypothesis 6 is valid in cases where profitability is proxied by ROA and NIM since INF has a significant impact on the VCB's profitability, but is invalid where profitability is proxied by ROE since the coefficient of INF is not different from zero. Regarding this, it can be said that the trend of Vietnam's inflation rate is dramatically decreasing in the sample period. The result of INF is positively related to VCBs' profitability, implying that during the period under this study, levels of inflation have been anticipated by VCBs. This enables them to adjust interest rates applied to loans and deposits accordingly, which consequently leads to higher profits (Driver & Windram, 2009).

**Hypothesis 7**: *The profitability of VCBs as measured by ROA, ROE and NIM is persistent over time.*

As observed in Table 8, the coefficient of the lagged ROA variable is positive but not significant. The result contradicts the findings of Athanasonglou *et al.* (2005), who show that ROA as a proxy for profitability tends to persist to a moderate extent, and Goddard *et al.* (2004), who suggest that the existence of profit in European banks is weak.

In addition, the positive coefficients of the one-period lagged ROE and NIM variables with p-values of 0.019 and 0.002, respectively, indicate that $H_0$ of the coefficients of zero are rejected at the 5% and 1% level, respectively. In other words, the one-period lagged ROE variable (ROE(-1)) is statistically significant at 10% and 5% levels, while lagged NIM is significant at 10%, 5%, and 1% levels. The significant coefficients of the lagged ROE and NIM confirm the dynamic nature of the model specification. ROE (-1) and NIM (-1) are positively and significantly related to ROE and NIM, meaning that banks with a high level of ROE and NIM in the previous year earn more ROA and NIM in the current year and vice versa. The regression results show that an increase of 1% in ROA and NIM from the previous year will increase current ROE and NIM by 0.48% and 1.28%, respectively. This result is consistent with the study conducted by Rahman (2015).

Therefore, we can conclude that hypothesis 7 is valid in the ROE and NIM models but is invalid in the ROA model. In other words, the profitability of VCBs is persistent when measured by ROE and NIM. Besides, Figure 2 below gives us a look at the trend in the profitability of VCBs. As can be seen, there has been a decreasing trend in the ROE of VBCs from 2006 to 2015. ROA and NIM are rather low compared to ROE. The reason could be an increase in competition in the banking industry in Vietnam. However, the persistence of VCBs' profitability as measured by ROE and NIM during the research period also shows us that VCBs have made an effort to obtain their profitability target.

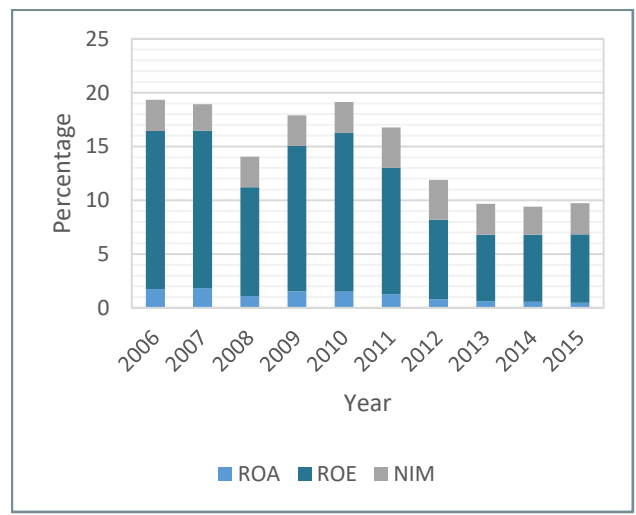

**Figure 2:** ROA, ROE, and NIM of VCBs (2006 – 2015). *Source*: VCBs data (Computed by the author)

## 6 Robustness check

To ensure robustness of results based on the dynamic Difference GMM method, it is important to test assumptions that satisfy Arellano & Bond (1991) one-step GMM dynamic panel estimator, which contributes to a firm base for our model and validate that the Difference GMM approach would actually yield valuable estimates of the coefficients of the regression that are more reliable.

1. **AR test:** Table 8 shows that, though the equations indicate that negative AR(1) (with relatively small p-values) is present, this does not imply that the estimates are inconsistent. Inconsistency will be implied if AR(2) is present and $H_0$ for AR(2) is rejected (Arellano & Bond, 1991). Also, as mentioned above, the AR(2) test is more important since it examines autocorrelation in levels, so we will use AR(2). The p-values of AR(2) in the ROA, ROE, and NIM models are 0.153, 0.101, and 0.330, respectively, which are greater than the 10% level, hence it is impossible to reject $H_0$. It indicates that three models have no autocorrelation in AR(2).

2. **Validity of Instrument Variables:** From Table 8, Sargan tests in all three models show no evidence of over-identifying restrictions. The Sargan test of the ROA model distributed $\chi^2$ (19) of 16.69 with a p-value of 0.611, and the Sargan test of the ROE and NIM models, distributed $\chi^2$ (19) of 14.13 with a p-value of 0.776 and $\chi^2$ (6) of 5.74 with a p-value of 0.45, indicate that $H_0$ cannot be rejected, meaning that the instrument variables are valid.

   On the other hand, when the number of instruments is greater than the number of commercial banks, the Sargan test may be weak; hence, according to the rule, *the number of instruments chosen should be less than or equal to the number of groups*. In this case, the number of instruments and the number of groups in the ROA and ROE models are equal to 26, while in the NIM model, the number of instruments (13) is half less than the number of groups (26). This says that in all three models, the rule of thumb for a number of instruments is kept. Thus, the results of the Sargan test are valid.

Finally, the results of the difference-in-Hansen tests of exogeneity of instrument subsets show p-values of difference are large (>10% level), implying that the difference instrument subsets are exogenous.

In sum, all assumptions about the dynamic Difference GMM are held, and the results of the dynamic Difference GMM estimator are reliable. In other words, the dynamic Difference GMM method is an entirely suitable choice for estimating the impact of credit risk on the financial performance of VCBs.

## 7 Conclusion and Implications

The study aims to seek empirical evidence to explain and discover how credit risk has a relationship with the profitability of VCBs. If so, what is the effect level? In order to discover this relation, the study uses an unbalanced panel data of 255 observations from 26 commercial banks covering the years 2006 - 2016. The study uses SIZE, NPLR, LLPR, CAR, GDP, and INF as credit risk indicators and ROA, ROE, and NIM as profitability indicators. The assumption tests show the models have autocorrelation, non-constant variance, and endogeneity. Hence, besides using regression methods including Pooled OLS, FEM, REM, and FGLS, dynamic Difference GMM is finally utilized to thoroughly address these problems. The study uses both basic and dynamic panel data; with basic panel data, POLS, FEM, REM, and FGLS are used to estimate the models, while with dynamic panel data, dynamic Difference GMM is employed to estimate the models. However, estimation results from POLS, FEM, and REM are not convincing enough since these methods should only be used when the standard assumptions are met (Molyneux *et al*., 2013). Once one of those assumptions, such as autocorrelation or heteroskedasticity, is violated, the estimation results are still unbiased and consistent but inefficient (t and F-tests are unreliable). The purpose of the estimation with these methods is to check the structural change of variables used in the models.

The results of the specification test indicate that there is a cross-section fixed effect and fixed period in the ROA and ROE models and a random effect in the NIM model. To overcome these defects, FGLS is used to estimate the models and then compared with the estimation results of ROA and ROE under FEM and NIM under REM so that we can see how the structural change of variables in models. The results are not too much different in comparison to these methods. However, the FGLS method requires the strict exogeneity of explanatory variables (Wooldridge, 2001), and we should be careful with a small sample size (Wooldridge, 2016, p. 384). Therefore, in order to check the robustness of the estimation results, the Difference GMM method, which can resolve the defects in the models (autocorrelation, heteroskedasticity, and endogeneity), is employed as final estimates. Then, based on the estimation results under this method, the study proceeds to analyze and discuss the models.

After building up the research objective, seven research hypotheses are formed. To test these hypotheses, the study has conducted regression analysis with the dynamic Difference GMM method. After estimating models with the Difference GMM estimator, in order to ensure unbiased, consistent, and efficient results, the assumptions of the dynamic Difference GMM estimator are tested. The test results show that both the Arrelano–Bond AR(2) tests for autocorrelation in residuals of order 2 and the Hansen test for over-identifying restrictions are highly significant, implying that all three models have no second-order autocorrelation and the instrument variables chosen are valid. In other words, all the Difference GMM estimator assumptions are satisfied, hence the estimation results under this method tend to be unbiased, consistent, and efficient.

The results of the regression under dynamic Difference GMM show that the high significance of the lagged dependent variable's coefficients in the ROE and NIM models confirm the dynamic character of the model specification, thus justifying the use of dynamic panel data model estimation. The results also show that despite the competition growth in Vietnamese financial markets, there is still a significant persistence of profit, measured by ROE and NIM, from one year to the next. It implies that if a bank makes an abnormal profit in the present year, then its expected profit for the following year will include a sizeable proportion of the present year's abnormal profit (Goddard *et al.*, 2004). The empirical findings also suggest that credit risk, proxied by GDP, has a significant positive impact on VCBs' profitability measured by ROA, ROE, and NIM. On the contrary, beyond our expectations, LLPR and CAR are not found to be important determinants of ROA, ROE, and NIM. It is also evident that the impact of SIZE, NPLR, and INF is not uniform across the different measures of VCBs' profitability used in this study. SIZE is found to be an important determinant of profitability measured by ROA and ROE but not by NIM. NPLR has a significant and negative impact on ROA and ROE, but the relationship is insignificant and negative in the NIM model. It is also evident that INF is positively and significantly related to ROA and NIM but is insignificantly related to ROE.

From the analysis and discussion of results, the study has some implications, including: From a regulatory perspective, VCBs should be based on individual commercial banks' efficiency. Policy on credit risk management should be enhanced to improve asset quality, hence minimizing non-performing loans, which tend to affect profitability and lead a bank to the banking crisis and an economy to a systematic crisis, and putting ROA, ROE, and NIM under control. Consequently, strong monitoring and control of assets should be done by both bank management and regulatory authorities.

## Acknowledgements

I would like to thank Dr. Tingting Zhu for her valuable comments and discussion. This paper is based on my thesis at Leicester University, where it was awarded the 2016 Best Full-Time Dissertation Prize.

**Appendices**

**Appendix 1**. List of Vietnamese commercial banks as of 20 May 2016

| No. | Abbreviation | Registered banks as of 20 May 2016 |
| --- | --- | --- |
| 1 | ABBANK | An Binh Commercial Joint Stock Bank |
| 2 | ACB | Asia Commercial Bank |
| 3 | AGRIBANK | Vietnam Bank for Agriculture and Rural Development |
| 4 | BIDV | Bank for Investment and Development of Vietnam |
| 5 | DONGABANK | Dong A Commercial Joint Stock Bank |
| 6 | EXIMBANK | Vietnam Commercial Joint Stock Export Import Bank |
| 7 | HDBANK | Ho Chi Minh Development Commercial Joint Stock Bank |
| 8 | KIENLONGBANK | Kien Long Commercial Joint Stock Bank |
| 9 | LIENVIETPOSTBANK | Lien Viet Post Joint Stock Commercial Bank |
| 10 | MB | Military Commercial Joint Stock Bank |
| 11 | MARITIME BANK | Vietnam Maritime Commercial Joint Stock Bank |
| 12 | NAMABANK | Nam A Commercial Joint Stock Bank |
| 13 | NCB | National Citizen Commercial Bank Joint Stock Bank |
| 14 | OCB | Orient Commercial Bank |
| 15 | SACOMBANK | Sai Gon Thuong Tin Commercial Joint Stock Bank |
| 16 | SAIGONBANK | Sai Gon Bank for Industry and Trade |
| 17 | SCB | Sai Gon Commercial Joint Stock Commercial Bank |
| 18 | SEABANK | Southeast Asia Commercial Joint Stock Bank |
| 19 | SHBANK | Sai Gon Ha Noi Commercial Joint Stock Bank |
| 20 | TECHCOMBANK | Vietnam Technological & Commercial Joint Stock Bank |
| 21 | VIB | Vietnam International Commercial Joint Stock Bank |
| 22 | VIETABANK | Vietnam Asia Commercial Joint Stock Bank |
| 23 | VIETCAPITALBANK | VietCapitalBank Commercial Joint Stock Bank |
| 24 | VIETCOMBANK | Bank for Foreign Trade of Vietnam |
| 25 | VIETINBANK | Vietnam Joint Stock Commercial Bank for Industry and Trade |
| 26 | VPBANK | Vietnam Prosperity Joint Stock Commercial Bank |

**Appendix 2.** Multicollinearity test for explanatory variables

| Variables | Variance Inflation Factor (VIF) | 1/ Variance Inflation Factor |
|---|---|---|
| SIZE | 1.85 | 0.540148 |
| NPLR | 1.45 | 0.687726 |
| LLPR | 1.74 | 0.574365 |
| CAR | 1.39 | 0.721413 |
| GDP | 1.24 | 0.808230 |
| INF | 1.10 | 0.913147 |
| Mean VIF | 1.46 | |

**Notes**: Multicollinearity exists in the model in case VIF > 10.0 (see Gujarati (2022, p. 340)). The VIF for the variables in the model ranges from 1.10 to 1.85 suggesting the absence of multicollinearity.

**Appendix 3.** Unit root test for all variables in the models

| Method | Inversed $\chi^2$ | | Inverse Normal | | Inverse logit | | Modified inv. $\chi^2$ | |
|---|---|---|---|---|---|---|---|---|
| Variables | Statistic | p-value | Statistic | p-value | Statistic | p-value | Statistic | p-value |
| ROA | 103.2616 | 0.0000 | -2.3687 | 0.0089 | -2.9773 | 0.0017 | 5.0266 | 0.0000 |
| ROE | 70.3873 | 0.0456 | -1.5715 | 0.0580 | -1.7455 | 0.0416 | 1.8030 | 0.0357 |
| NIM | 78.6911 | 0.0099 | -2.6736 | 0.0038 | -2.7345 | 0.0035 | 2.6173 | 0.0044 |
| SIZE | 302.2862 | 0.0000 | -9.8132 | 0.0000 | -14.9759 | 0.0000 | 24.5426 | 0.0000 |
| NPLR | 78.5956 | 0.0100 | -2.4419 | 0.0073 | -2.4744 | 0.0073 | 2.6079 | 0.0046 |
| LLPR | 85.1055 | 0.0026 | -1.9168 | 0.0276 | -2.6560 | 0.0044 | 3.2463 | 0.0006 |
| CAR | 152.9311 | 0.0000 | -4.1305 | 0.0000 | -6.7480 | 0.0000 | 9.8971 | 0.0000 |
| GDP | 77.2393 | 0.0131 | -3.7303 | 0.0001 | -6.7480 | 0.0004 | 2.4749 | 0.0067 |
| INF | 88.7773 | 0.0011 | -4.4953 | 0.0000 | -4.1454 | 0.0000 | 3.6063 | 0.0002 |
| Test Result | Stationary | | | | | | | |

**Notes**: Under F-test (Choice, 2001), H₀ is all the panels containing unit roots (i.e. variables of time-series are non-stationary), including four methods: the inverse $\chi^2$, inverse normal, and inverse logit transformations. Though H₀ of ROE is not rejected (p-value of 0.0580) under the Inverse Normal method, the number of panels in the model is finite (255 panels), and the inverse $\chi^2$ p-test is applicable. $\chi^2$ p-value of ROE is 0.0456 which is less than a 5% level allowing us to reject H₀, so variables in the three models are stationary.

**Appendix 4**. Endogeneity and exogeneity test result summary

| Variable | | | SIZE | NPLR | LLPR | CAR | GDP | INF | Lags |
|---|---|---|---|---|---|---|---|---|---|
| ROA | D | Chi2(1) | 1.02419 | 2.01823 | 35.6111 | .613042 | 1.50355 | 3.10084 | .11155 |
| | | p-value | 0.3115 | 0.1554 | 0.0000 | 0.4336 | 0.2201 | 0.0783 | 0.7384 |
| | W | F-test | 1.01215 | 2.00236 | 40.7658 | .604847 | 1.48869 | 3.08972 | .109842 |
| | | p-value | 0.3154 | 0.1583 | 0.0000 | 0.4375 | 0.2236 | 0.0800 | 0.7406 |
| Results | | | Exo. | Exo. | Endo. | Exo. | Exo. | Exo | Exo. |
| ROE | D | Chi2(1) | .613126 | 2.38506 | 4.09748 | .119181 | .089492 | 3.13512 | 2.0057 |
| | | p-value | 0.4336 | 0.1225 | 0.0429 | 0.7299 | 0.7648 | 0.0766 | 0.1567 |
| | W | F-test | .604931 | 2.36976 | 4.09908 | .117359 | .088113 | 3.12431 | 1.98982 |
| | | p-value | 0.4374 | 0.1250 | 0.0440 | 0.7322 | 0.7668 | 0.0784 | 0.1596 |
| Results | | | Exo. | Exo. | Endo. | Exo. | Exo. | Endo. | Exo. |
| NIM | D | Chi2(1) | .10764 | 2.77511 | 26.2493 | .147299 | 2.41059 | .123546 | .174199 |
| | | p-value | 0.7428 | 0.0957 | 0.0000 | 0.7011 | 0.1205 | 0.7252 | 0.6764 |
| | W | F-test | .10599 | 2.76158 | .46735 | .145064 | 2.39536 | .121659 | .171573 |
| | | p-value | 0.7450 | 0.0978 | 0.0000 | 0.7036 | 0.1230 | 0.7275 | 0.6791 |
| Results | | | Exo. | Exo. | Endo. | Exo. | Exo. | Exo. | Exo. |

**Notes:** Under Durbin (D) and Wu-Hausman (W) test, the null hypothesis ($H_0$) denotes exogenous (Exo) variables, otherwise alternative ($H_1$) is endogenous (Endo) variables. Lags present lags of ROA, ROE, and NIM respectively.